\documentclass[12pt,preprint]{aastex}
\shorttitle{Hot Dust and PAH Emission at Low Metallicity}
\shortauthors{Jackson et al.}

\begin{document}

\def\HI{\ion{H}{1}}
\def\HII{\ion{H}{2}}
\def\msun{M_\sun}
\def\spitzer{$\it Spitzer$}
\def\micron{$\mu$m}

\title{Hot Dust and PAH Emission at Low Metallicity: A Spitzer Survey of Local Group and Other Nearby Dwarf Galaxies}

\author{Dale C. Jackson}
\affil{Astronomy Department, University of Minnesota, 
116 Church St S.E., Minneapolis, MN 55455}
\email{djackson@astro.umn.edu}

\author{John M. Cannon}
\affil{Max Planck Institut Astronomic, K\"{o}nigstuhl 17, 
D-69117 Heidelberg, Germany}
\email{cannon@mpia.de}

\and

\author{Evan D. Skillman, Henry Lee, Robert D. Gehrz, 
Charles E. Woodward, and Elisha Polomski}
\affil{Astronomy Department, University of Minnesota, 
116 Church St S.E., Minneapolis, MN 55455}
\email{skillman@astro.umn.edu, hlee@astro.umn.edu,
gehrz@astro.umn.edu, chelsea@astro.umn.edu, elwood@astro.umn.edu}

\begin{abstract}
We present \spitzer\ Space Telescope 4.5 and 8.0 \micron\ imaging of
15 Local Group and nearby dwarf galaxies. We find that the diffuse 8
micron emission is spatially correlated with regions of active star
formation in these spatially resolved galaxies. Our sample spans a
range of more than one dex in nebular metallicity and over three
orders of magnitude in current star formation rate, allowing us to
examine the dependence of the diffuse 8 \micron\ emission, originating
from hot dust and PAHs, on these parameters. We detect prominent
diffuse 8 \micron\ emission in four of the most luminous galaxies in
the sample (IC~1613, IC~5152, NGC~55, and NGC~3109), low surface
brightness emission from four others (DDO~216, Sextans~A, Sextans~B,
WLM), and no diffuse emission from the remaining objects.  These data
are the first spatially resolved images of diffuse 8 \micron\ emission
from such low-metallicity objects (12+log(O/H)$\sim$7.5). We observe
general correlations of the diffuse 8 \micron\ emission with both the
current star formation rate and the nebular metallicity of the
galaxies in our sample. However, we also see exceptions to these
correlations that suggest other processes may also have a significant
effect on the generation of hot dust/PAH emission. These systems all
have evidence for old and intermediate age star formation, thus the
lack of diffuse 8 \micron\ emission cannot be attributed to young
galaxy ages.  Also, we find that winds are unlikely to explain the
paucity of diffuse 8 \micron\ emission, since high resolution imaging
of the neutral gas in these objects show no evidence of blowout.
Additionally, we propose the lack of diffuse 8 \micron\ emission in
low-metallicity systems may be due to the destruction of dust grains
by supernova shocks, assuming the timescale to regrow dust grains and
PAH molecules is long compared to the destruction timescale.  The most
likely explanation for the observed weak diffuse 8 \micron\ emission
is at least partly due to a general absence of dust (including PAHs),
in agreement with their low metallicities.
\end{abstract}


%%%%%%%%%%%%%%%%%%%%%%%%%%%%%%%%%%%%%%%%%%%%%%%%%%%%%%%%%%%%%%%%%%%%%%

\section{Introduction}
Nearby galaxies have proven to be important laboratories for the study
of the basic properties and evolutionary histories of all
galaxies. This is because we can study both their stellar populations
and ISM at high spatial resolution. While the stellar and gaseous
contents of Local Group dwarf galaxies have been studied in depth
\citep[][and references therein]{mat98,vdb00}, their 5-200 micron
thermal dust continuum emission has only been examined in detail in a
few local systems, including the Magellanic Clouds
\citep{con98,rea00,con00,stu00,ver02} and NGC 6822 \citep[][in
preparation]{gal91,isr96,hun01,can06}. The nature of far-infrared
emission in other Local Group dwarf galaxies has primarily been
inferred from spatially unresolved IRAS observations (e.g.,
\citealt{hun89,mel94}), due in large part to our previous inability to
perform sensitive wide-field, high-resolution imaging at these
wavelengths.

Of particular interest in these infrared images is the emission
attributed to polycyclic aromatic hydrocarbons (PAHs). Emission from
these macro-molecules has been detected in numerous massive galaxies,
often highlighting the photodissociation regions (PDRs) surrounding
large \HII\ regions \citep{gia94, hel04}.  This morphological
coincidence is normally attributed to the hard radiation field inside
the \HII\ region dissociating the molecules \citep{bou90, hel01},
while outside the PDR there are insufficient UV photons to excite PAHs
into emission. However, PAHs have also been found in the quiescent
ISM, as reported by \citet{lem98} in their detection of PAHs in an
isolated cirrus cloud illuminated only by the typical interstellar
radiation field of the solar neighborhood.

What is not well understood is the dust and PAH content in lower mass,
lower metallicity systems. It is assumed that these systems will
exhibit less PAH emission due to the ability of the interstellar
radiation field to penetrate through lower dust column densities (due
to potentially lower dust-to-gas ratios in low-metallicity
environments) dissociating or destroying the PAHs
\citep{gal03}. However, sensitivity has limited previous mid-infrared
studies to massive galaxies and starburst systems
\citep*[e.g.,][]{gal03, gal05, mad05}. These types of studies have
also been performed with \spitzer\ \citep{eng05,hog05,wu05,oha05},
including studies of lower mass, lower luminosity objects
\citep{wu05,ros06}. However, for quiescent dwarf irregular galaxies
the dependence of hot dust/PAH emission on galaxy properties such as
metallicity or current star formation rate remains relatively
unexplored. The unprecedented sensitivity and spatial resolution of
\spitzer\ for the first time enables us to directly image the spatial
distribution of the hot dust/PAH population in these low surface
brightness systems. In addition, the close proximity of these galaxies
allows us to discern the diffuse dust emission from the underlying
stellar population, which is still not possible in more distant
systems.

In this paper we present \spitzer/Infrared Array Camera (IRAC) 4.5 and
8.0 \micron\ images of 15 Local Group and nearby dwarf galaxies. The
continuum subtracted 8 \micron\ images are sensitive to both the 7.7
\micron\ PAH feature as well as any hot ($\sim$350 K) dust present in
these objects. Our galaxy sample spans a range of over one dex in
metallicity, 8.5 mag in absolute B magnitude, and three orders of
magnitude in current star formation rate to which we can compare the
diffuse 8 \micron\ flux. In \S \ref{observations} we describe the
observations, data reduction, and continuum subtraction procedures. In
\S \ref{individual} we present our images and discuss interesting
features of individual objects. In \S \ref{discussion} we investigate
correlations of diffuse 8 \micron\ flux with other galaxy properties
and compare these results with similar studies and discuss possible
causes for the drop of diffuse 8 \micron\ flux at low metallicity. We
summarize our findings in \S \ref{conclusions}.

%%%%%%%%%%%%%%%%%%%%%%%%%%%%%%%%%%%%%%%%%%%%%%%%%%%%%%%%%%%%%%%%%%%%%%%%%%%%

\section{Observations and Data Reduction}\label{observations}
Our galaxy sample was chosen from the list of all dwarf galaxies
judged by Mateo to be potential Local Group members (as reported in
\citealt[][Table 1]{mat98}) based on heliocentric radial velocities
and an estimate of the Local Group zero-velocity surface.  We have
selected all the irregular and transition (dIrr/dSph) galaxies that
are more than 20$\degr$ above or below the galactic plane and that
were not already listed in the \spitzer\ Reserved Object Catalog.

The data were acquired with the
IRAC camera \citep{faz04} on the \spitzer\ Space Telescope
\citep{wer04} as part of a guaranteed time program (PID \#128,
P.I. R.D. Gehrz) that consists of imaging and spectroscopy of 16 Local
Group and nearby dwarf galaxies. While IRAC is capable of imaging at
3.6, 4.5, 5.8, and 8.0 \micron\ (hereafter channels 1-4 respectively),
the field of view of channels 1 and 3 is offset from that of channels
2 and 4. In this exploratory program, we chose to image our targets in
only channels 2 and 4, since these bandpasses allow us to detect both
the stellar population (channel 2) and the hot dust/PAH (channel 4)
components simultaneously. For all targets in both channels, we
performed five dither positions with exposure times of 200 seconds
each, except for NGC~55 and WLM which had exposure times of 100
seconds per dither position. A single 5$\arcmin$ $\times$ 5$\arcmin$
pointing was used for all targets except WLM, NGC~55, NGC~3109, and
IC~1613. WLM, NGC 3109, and IC~1613 were mosaicked in array
coordinates using 1 $\times$ 3, 1 $\times$ 2, and 2 $\times$ 2
pointing grids respectively (see Table \ref{sample}). NGC~55 was
mosaicked in sky coordinates using a 1 $\times$ 3 pointing mosaic.

The data were reduced using the MOPEX\footnote{MOPEX is available from
the Spitzer Science Center at http://ssc.spitzer.caltech.edu/postbcd/}
reduction package version 2004 October 15. Detector artifacts such as
column pull-down and muxbleed were corrected with the `cosmetic'
program. These corrected data were then used as the input images for
the `overlap' routine which matches the backgrounds of individual
frames to create a smooth background in the final mosaic. Outlier
detection, image interpolation, and image co-addition were done with
the `mosaic' task; this program also produced the final mosaics with a
pixel scale of 0.86\arcsec , which was motivated by
\citet{van00}. Some of our channel 4 data were strongly affected by
persistent images resulting from bright sources imaged in the program
immediately before our data were taken.  Subtracting a
background-subtracted median-combined image from each frame proved
very effective at removing these artifacts.

We then created the `continuum-subtracted' images. While a single
fiducial image frame was used to create all four IRAC mosaics, the
image alignment was still not adequate. An additional translation for
each galaxy (typically 0.5~-~1.2 pixels in x and y) was required to
align the channel 2 and channel 4 images. The channel 2 image was
aligned with the channel 4 image by calculating the centroids of point
sources in both channel 2 and channel 4 images, and shifting the
channel 2 image to align it with the coordinates measured in channel
4. The channel 2 image was then convolved with a PSF
kernel\footnote{The PSF convolution kernel was kindly provided by
K. Gordon and is publicly available at
\url{http://dirty.as.arizona.edu/$\sim$kgordon/mips/conv$\_$psfs/conv$\_$psfs.html}}
to attempt to reproduce the channel 4 PSF. Extensive tests were
performed, including aperture photometry on the residual `diffuse'
image. We found no evidence for systematic differences between the
modeled and observed channel 4 PSF compared to the overall flux
uncertainty in the images.  Next, we subtracted the background value
from the convolved channel 2 image, which we determined by taking the
mode of non-crowded regions in the image. And finally, we multiplied
the channel 2 image by a scale factor (listed in Table \ref{data}) and
subtracted it from the channel 4 image to study the diffuse emission
at 8 \micron .

This scale factor was determined empirically by finding the value that
removed the majority of the underlying point sources without leaving
residual peaks or holes. This process is inherently uncertain since
adopting one color for the entire stellar population is a rather poor
approximation. The errors associated with this method are also
exaggerated because we are typically dealing with a few tens to
hundreds of resolved point sources, as opposed to the smooth stellar
population in unresolved galaxies. We find in general that the best
value for the scale factor was 0.4, which is consistent with the
studies of \citet{hel04} and \citet{pah04}. This consistency is
notable, since in our low-metallicity systems the ratio of M-type AGB
stars to carbon stars is significantly higher than in more metal-rich
systems.  Because AGB stars constitute the majority of the point
sources we detect in channel 4, this would lead us to expect, on
average, redder sources than in the more metal-rich sources studies by
the former authors. Indeed, the average channel 2-4 color of point
sources for the galaxy WLM is 1.07 (Jackson et al. 2006, in
preparation), yet applying the scale factor that corresponds to this
color (0.95) produces an extreme over-subtraction for a large fraction
of the sources and a negative total diffuse 8 \micron\ total flux for
the galaxy (please note, throughout this work all magnitudes are
Vega-relative). Further, we find a different scale factor for NGC~55
(0.7), which is the most metal-rich object in our sample.

For all of the galaxies in our sample we adopt a 3$\sigma$ scale
factor error of 50\%. This value was chosen because tests showed
adopting a scale factor 50\% higher would result in a negative total
diffuse 8 \micron\ flux for the four galaxies in our sample with the
lowest current star formation rate, and therefore the least expected
diffuse 8 \micron\ emission.  The total flux uncertainty in the
channel 2 and channel 4 images of all the galaxies is assumed to be
10\% \citep{hor04}.  The photometric uncertainty we report for the
diffuse 8 \micron\ fluxes were calculated using the error propagation
equation, which included the total flux uncertainties from the channel
2 and channel 4 images and the uncertainty in the scale factor.  We
include no independent error due to foreground star contamination,
since foreground stars are treated in the same manner as stars in
their host galaxy in the continuum subtraction process and are
therefore included in the scale factor error estimation.

The images were trimmed to remove the low signal-to-noise edges that
result from the dithering pattern we chose. Otherwise the entire
images were used to derive the fluxes except in the case of Sextans~A,
which has an extremely bright foreground star in the field. The flux
calculation for this object is described in \S\ref{sextans_a}.  For
all of the objects except Sextans~A, we calculated the diffuse 8
\micron\ flux by finding the sum of the residual flux with `imstat' in
IRAF\footnote{IRAF is distributed by the National Optical Astronomy
Observatories, which are operated by the Association of Universities
for Research in Astronomy, Inc., under cooperative agreement with the
National Science Foundation.}, applying the correction for an infinite
aperture, and subtracting the total background.

As an example of the complexity in interpreting diffuse 8 emission in
galaxies, \citet{can06} have found spatially variable PAH emission
strengths in the dwarf starburst galaxy NGC 1705; mid-IR spectra of
two regions, with the same nebular metallicity, show very different
PAH to continuum ratios. Since we are unable to distinguish between
PAH emission and thermal continuum from hot dust in our broadband
images, we make no color-correction in calculating the 8 \micron\
fluxes. Also because of the ambiguity between PAH and thermal dust
continuum emission, we will refer throughout this paper to the sum of
these components as `diffuse 8 \micron\ emission.' Future \spitzer\
IRS spectra from regions of bright 8 \micron\ emission in these
objects will be necessary to distinguish between PAH and hot dust
emission (e.g., \citealt{hou04,smi04}).

%%%%%%%%%%%%%%%%%%%%%%%%%%%%%%%%%%%%%%%%%%%%%%%%%%%%%%%%%%%%%%%%%%%%%%%%%%%

\section{Individual Objects}\label{individual}
In the following sections we briefly describe the relationships
between diffuse 8 \micron\ emission (for the objects we detect) and
observations at other wavelengths. Images of the galaxies in which we
detect prominent diffuse 8 \micron\ emission are shown in Figure
\ref{n55} (for NGC~55) and Figure \ref{3detect} (for IC~5152,
NGC~3109, and IC~1613). Objects with only low surface brightness
emission are shown in Figure \ref{split1} (for WLM and Sextans A) and
Figure \ref{split2} (for Sextans B and DDO~216). The objects with no
detected diffuse 8 \micron\ emission (Antlia, DDO~210, GR~8, Leo~A,
LGS~3, Phoenix dwarf irregular, and UGCA~438) are shown in Figure
\ref{nodetect}.

\subsection{DDO 216 (Peg DIG)}
This low-luminosity galaxy has very weak H$\alpha$ emission, as there
are only one compact H~II region and one diffuse H~II region known
\citep{hun93,ag95,skillman97}.  Most of the diffuse 8~\micron\
emission in the dwarf galaxy is concentrated about the two H~II
regions, coincident with the peak of the H~I distribution
\citep{young03}.  The measured present-day star formation rate is very
small ($\la 10^{-4}$ $M_{\odot}$~yr$^{-1}$; \citealp{young03}), and is
much lower than the mean star formation rate over the lifetime of the
galaxy.  With {\em HST\/} stellar photometry, \citet{gallagher98}
showed that the star formation rate was elevated by a factor of a few
about $\sim$ 1~Gyr ago compared to the present day star formation
rate.

\subsection{IC~1613}
\cite{skillman03} obtained deep {\em HST\/} photometry of stars in an
outer field of the galaxy, and demonstrated that the star-formation
rate $\sim$ 3--5 Gyr ago was enhanced by roughly a factor of two
compared to the present-day star-formation rate.  Most of the diffuse
8~\micron\ emission is roughly coincident with the complex of
H$\alpha$ bubble-, loop-, and filamentary-structure in the northeast
part of the galaxy (see \citealt{hod90,hun93}).  The H$\alpha$ complex
contains a well-studied supernova remnant (SNR; \citealp{dodb80}),
labeled H~II region \#49 (\citealp{hod90}; also known as Sandage 8 or
S8). This SNR has been studied recently by \cite{vg01} and
\cite{rosado01}, the latter of which showed that the superbubbles have
dynamical ages of $\sim$ 1--2~Myr.  They also suggested that the SNR
is partly hidden by dust, because the SNR is located near the
intersection of two to three large H$\alpha$ shells (R4, R9, and R10
in \citealp{vg01}), in which there appears to be material with high
obscuration.  In fact, the peak of the 8~\micron\ emission appears to
be located between the SNR and an H~II region just to the northwest
(H~II \#39 in \citealp{hod90}).  The location is near the east-west
midpoint of the obscuration ridge seen in H$\alpha$ velocity maps by
\cite{vg01}.  It is not surprising that most of the 8~\micron\
emission is found near the locations of the H$\alpha$ bubbles, where
there may be a concentration of molecular and dusty material.  We
should point out that two additional very faint patches of diffuse
8~\micron\ emission are coincident with the positions of the bright
H~II regions \#37 near the southeast corner of the galaxy and \#13
towards the western side of the galaxy.

\subsection{IC~5152}
Despite the relative isolation, modest luminosity, and relatively low
gas-phase metallicity, this dwarf contains regions of bright
8~\micron\ emission which lie in the mid-plane of the galaxy.
Unfortunately, there is no published H$\alpha$ map available for
comparison.  However, a bright 8~\micron\ peak on the eastern side
appears to correspond to H~II region ``A,'' whose optical spectrum was
measured by \cite{talent80}, \cite{webster83}, and \cite{lee03south}.
The diffuse 8 \micron\ peak near the center of the galaxy appears to
correspond to a large grouping of bright ultraviolet point sources
\citep{maoz96}, which are likely O- and B-type stars; their
corresponding optical sources (though unresolved) were noted by
\cite{zm99}.

\subsection{NGC~55}
NGC~55 is a Magellanic spiral galaxy, and is either a member of the
Local Group \citep{mat98} or the Sculptor Group \citep{vdb00}. The
near edge-on orientation \citep[i~=~83$\degr$;][]{kar04} makes this
galaxy an excellent candidate to examine the role of disk H~II regions
in extraplanar emission.  Published H$\alpha$ imaging has shown an
array of structure (arcs, loops, filaments, etc.) in the extraplanar
diffuse ionized gas (e.g., \citealp{ferguson96,hoopes96,od99,mv03}).
The variety of structure and intensity seen in diffuse 8~\micron\
emission appears to match similarly with those in H$\alpha$. For
example, the shell-like structure to the northeast in 8~\micron\ at
$\alpha,\delta$ (J2000) = $00^h15^m12^s, -39\degr12\arcmin35\arcsec$
corresponds in H$\alpha$ with the U-shaped structure and H~II region
H5 in \cite{od99}.  This shell, another similar shell immediately to
the south, and the central emission regions correspond to similar
features seen at 24~\micron\ \citep{eng04}.  There is also a large
faint shell in 8~\micron\ (centered roughly at $00^h14^m46^s,
-39\degr12\arcmin08\arcsec$) just southwest of the center (``below''
the midplane) of the galaxy. This feature is seen as a chain of H~II
regions and bubbles in \cite{tuellmann03}.  One of the two extraplanar
H~II regions (EHR1, EHR2; \citealp{tuellmann03}) is barely visible in
8~\micron.  The measured flux of EHR1 in a circular aperture of radius
5\arcsec\ is $F$(8~\micron) = $90 \pm 10$~$\mu$Jy; EHR2 is not visible
in the image.

\subsection{NGC~3109}
Located just outside of the Local Group, NGC~3109 is an almost edge-on
\citep[i~=~86$\degr$;][]{kar04} Magellanic-type irregular galaxy, and
is the largest member of the very loose Antlia-Sextans group
\citep{vdb99,tully02}.  From ground-based stellar photometry,
\cite{alonso99} determined that a majority of luminous stars observed
in the near-infrared were likely $\lesssim$ 1~Gyr old.
\cite{minniti99} reported a halo of old, metal-poor stars, although
the carbon-star study by \cite{demers03} disputes the existence of
such a halo.  In addition to the existence of a large H~I halo, a
small H~I warp was discovered to the southwest, which may have been
caused by an interaction $\sim$ 1~Gyr ago with the neighboring Antlia
dwarf galaxy \citep{jc90,bb01}.  Only upper limits or non-detections
have been obtained for the molecular content from CO measurements
\citep{rr80,bresolin93}.  Local peaks in the diffuse 8~\micron\
emission are roughly coincident with the H~II regions (e.g.,
\citealp{rm92,bresolin93,hun93}).  The bright 8~\micron\ peak at
$\alpha,\delta$ (J2000) = $10^h03^m04^s, -26\degr09\arcmin22\arcsec$
corresponds to the H~II region \#6 in \cite{rm92} (D2H5 in
\citealp{bresolin93}).  The emission ridge to the northwest of the
aforementioned peak is likely associated with the northern edge of or
coincident with H~II regions \#5, \#4, and \#2 in \cite{rm92} (F1H3,
F1H1, and F1H2 in \citealp{bresolin93}).  The ``frothy'' infrared
emission to the east is related to the diffuse H$\alpha$ emission
\citep{bresolin93,hun93}.

\subsection{Sextans~A}\label{sextans_a}
Sextans~A is a dwarf irregular galaxy in the Antlia-Sextans group.
With {\em HST\/} photometry of the resolved stellar population,
\cite{dohmpalmer97} and \cite{dohmpalmer02} showed that (1) a region
with star-formation activity in the last 50~Myr coincided with the
brightest H~II regions and the peak in the H~I distribution
\citep{skillman88}, and (2) that the pattern of star formation was
occurring in a stochastic manner throughout the galaxy.  The most
recent star formation activity was shown to be about a factor of ten
higher than the average star-formation activity over a Hubble time.
\cite{rr80} obtained an upper limit for the CO molecular gas. The
extremely bright foreground star along the line of sight to Sextans~A
causes significant problems in our \spitzer\ data, since the continuum
subtraction of this object leaves broad negative wings across the
image. However, we detect low level diffuse 8~\micron\ emission
coincident with the bright H~II region \#17 \citep{hodge94} in the
southeastern corner of the galaxy. To minimize the effect of the
foreground star, we only summed the diffuse 8 \micron\ flux from this
region.  Consequently, the measured flux is plotted as a lower limit
in Figures \ref{3corr} and \ref{3groups}, since we are not sensitive
to any emission outside this region.

\subsection{Sextans~B}
Sextans~B is a another dwarf irregular galaxy in the Antlia-Sextans
group, and has properties closely resembling that of Sextans~A.
However, \cite{kniazev05} argue that there may be a spatial
inhomogeneity in the H~II region oxygen abundance by 0.3~dex over a
length scale of about 0.6~kpc.  The low surface brightness emission at
8~\micron\ appears to coincide with H~II regions \#1, \#5, and \#10
from \cite{strobel91}.  As in Sextans~A, there is only an upper limit
for the CO molecular gas \citep{rr80}.

\subsection{WLM}
\cite{hm95} derived a total present-day star formation rate (based on
the H$\alpha$ flux) of about $10^{-3}$ $M_{\odot}$~yr$^{-1}$, and
\cite{tk01} presented low upper limits to the total amount of CO gas
in the galaxy. {\em HST\/} photometry of the resolved stellar
populations \citep{dolphin00} suggests that increased star-formation,
mostly in the bar of the galaxy, has taken place recently in the last
$\simeq$ 1~Gyr.  Diffuse 8~\micron\ emission is mostly associated with
the two dominant H~II regions HM~7 and HM~9 (\citealp{hm95}; see also
\citealp{hun93,lsv05}).  \cite{jac04} and \cite{lsv05} suggested that
there could be excess dust or regions of higher extinction near the
second H~I peak in the southeast portion of the galaxy, which may
explain the small compact regions of 8~\micron\ emission.  However,
the spectroscopic line-of-sight reddening values (near zero) from the
latter author derived from the Balmer decrement for the bright H~II
regions HM~7 and HM~9 cannot explain the presence of 8~\micron\
emission.

%%%%%%%%%%%%%%%%%%%%%%%%%%%%%%%%%%%%%%%%%%%%%%%%%%%%%%%%%%%%%%%%%%%%%%%%%%%%%%

\section{Discussion}\label{discussion}
\subsection{Correlations of Diffuse 8 \micron\ Emission with Galaxy Properties}
ISO observations revealed, for the first time, the scarcity of PAH
emission in low-metallicity systems \citep{mad00}. More recently
though, similar studies with \spitzer\ have been carried out by
\citet{eng05}, and \citet{ros06}, among others. The observations
presented here are different in that the galaxies in our sample are
well-resolved, which allows us to address the spatial distribution of
the diffuse 8 \micron\ emission in these galaxies.  It is important to
note that we are required to treat our data differently than the
studies mentioned above. This is because the highly resolved nature of
our images allows us to determine that a substantial fraction of the 8
\micron\ emission in our images comes from the stellar population
(both foreground and in the galaxies themselves). Consequently, we
found it necessary to subtract the continuum from the images prior to
summing the 8 \micron\ flux. The studies of \citet{eng05} and
\citet{ros06} each address the expected stellar flux at 8.0 \micron\
and conclude no continuum subtraction is necessary for their samples.

In Table \ref{data} we list the diffuse 8 \micron\ fluxes along with
other galaxy properties. Of the 15 targets observed for this program
only NGC~55, IC~5152, IC~1613, and NGC~3109 have prominent diffuse
emission at 8 \micron, while DDO~216, Sextans~A, Sextans~B, and WLM
show only low surface brightness emission. Both \citet{eng05} and
\citet{ros06} have shown that there are correlations of the diffuse 8
\micron\ emission with both host galaxy metallicity and current star
formation rate. We will investigate the physical processes that are
responsible for these correlations later in this section. First,
however, we wish to examine the null hypothesis, which is that the
lack of diffuse 8 \micron\ emission in galaxies with low metallicity
and low star formation rate is simply an artifact of the correlations
of metallicity and star formation rate with galaxy mass. That is
to say, are the systems which seem to be deficient in diffuse 8
\micron\ emission simply smaller systems that we would expect to have
less total ISM, including PAHs and dust? Figure \ref{lumin} shows the
diffuse 8 \micron\ luminosity scaled by the total stellar mass as a
function of host galaxy metallicity (a) and current star formation
rate (b). The total stellar mass was inferred from the 4.5 \micron\
luminosity as described in \citet{lee06}, and the adopted distances to
the galaxies in our sample are from \citet{mat98}. The diffuse 8
\micron\ fluxes for our sample are plotted in all figures regardless
of if diffuse 8 \micron\ emission is obvious in the images. This
figure shows that lower metallicity systems do in fact have
disproportionately less diffuse 8 \micron\ than more metal-rich
objects.

Figure \ref{3corr} shows the diffuse 8 \micron\ absolute magnitudes
versus absolute B magnitude, nebular metallicity, and current star
formation rate (SFR) as measured by their H$\alpha$ fluxes [adopting
distances and SFRs from \citet{mat98} and a channel 4 magnitude zero
point of 64.1 Jy; \citealt{IRAC}] for each target in our sample. All
three quantities are correlated with the diffuse 8 \micron\ flux. In
Figure \ref{3groups} we plot the absolute diffuse 8 \micron\ magnitude
versus oxygen abundance and current star formation rate, along with
the data of \citet{ros06} and \citet{eng05}.

To date, the outstanding question has been whether it is the
decreasing metal content or the decreasing number of high energy
photons capable of heating/exciting the dust and PAH molecules into
emission that is responsible for the decrease in hot dust/PAH
emission. Spatially resolved observations here indicate that both of
these effects are important. For example, IC~1613 and WLM have very
similar gross properties. Despite these systems having the same
current star formation rate, IC~1613 shows prominent diffuse 8
\micron\ emission, while WLM has only low surface brightness
emission. This is probably due to the fact that the current star
formation is spread over a large area in WLM, while it is concentrated
in a single \HII\ region in IC~1613. This points to the fact that even
though these objects have the same SFR, the density of UV radiation in
the star forming region of IC~1613 must be higher than in those of
WLM. Thus, we observe prominent diffuse 8 \micron\ emission in IC~1613
and not in WLM, even though the nebular metallicity of WLM is 0.2 dex
higher.

In comparison, Sextans~A appears to have even weaker diffuse 8
\micron\ emission than WLM, although the foreground star in Sextans~A
makes a numerical comparison impossible. Again, these objects have
nearly the same gross properties (Sextans~A has a slightly lower SFR,
but the distribution of \HII\ regions are similar), however, Sextans~A
has a significantly lower metallicity, which is likely the cause of
the difference in diffuse 8 \micron\ emission between these two
objects.

\subsection{Physical Causes of the Lack of Diffuse 8 \micron\ Emission}
There is evidence to suggest that there may be more factors involved
than just the radiation field and metallicity of these
objects. \citet{lis98} found dust-to-gas ratios in dwarf galaxies to
vary by as much as a factor of 10 {\it at a given metallicity}. While
we do not want to over-interpret dust mass estimates derived from a
single-temperature blackbody fit to two data points, their data
suggest a large variation in the dust content between galaxies with
very similar gross characteristics.  Perhaps it should come as no
surprise then, that we also see a large variation in the hot dust/PAH
emission.

Still, the issue of what is responsible for the general drop in hot
dust/PAH emission in low-metallicity systems remains. \citet{dal01}
and \citet{hou04} performed mid-IR spectroscopy of the very
low-metallicity galaxy SBS 0335-052. \citet{dal01} conclude that PAH
emission is not seen in this galaxy because it is extremely young and
the stars that will eventually become AGB stars and enrich the ISM
with carbon have not yet evolved. The same explanation for other
galaxies was offered by \citet{eng05}, \citet{mad05}, and
\citet{oha05}. While a young galaxy age may explain the lack of PAH
emission in SBS 0335-052 and some other systems, this cannot be true
for the galaxies in our sample.  Since all of the objects in our
sample formed the bulk of their stars over 2 Gyr ago \citep{mat98},
the AGB has been well populated and therefore the lack of AGB stars
cannot be responsible for the absence of PAH emission in Local Group
dwarfs.

A possibility proposed by \citet{hog05} for the dearth of hot dust/PAH
emission in low metallicity systems is that the gravitational
potential of these systems is not sufficiently large to retain dust
and PAH molecules against the effects of vigorous star formation. We
are able to address this possibility directly, since high-resolution
\HI\ imaging has been performed for the majority of the galaxies in
our sample.

High-resolution \HI\ imaging has been published for 10 of the 15
targets in our sample [\citet{lo93} for DDO~210, DDO~216, Leo~A,
LGS~3, and GR~8; \citet{lak89} for IC~1613; \citet{you97} for LGS~3
and Phoenix; \citet{jc90} for NGC~3109; \citet{puc91} for NGC~55;
\citet{wil02} for Sextans~A] and lower-resolution imaging has been
done on two additional galaxies [WLM \citep{jac04} and Antlia
\citep{jc90}]. It has been established by the high-resolution studies
that the \HI\ distribution of irregular galaxies tends to be clumpy
and irregular in nature, with these clumps not spatially correlated
particularly well with the stellar populations of these systems. Also,
two of the largest, most massive systems in our sample (NGC~55 and
NGC~3109) both exhibit warps in their \HI\ disks, which have been
attributed to interactions with other nearby dwarf galaxies. In only
two of the galaxies in our sample is there any evidence that the
stellar population is responsible for expelling the ambient
ISM. \citet{jac04} found a double-peaked structure in the \HI\
distribution of WLM, which they believed might be due to a partial
blowout, but high-resolution \HI\ imaging is required to confirm this
result. \citet{you97} and \citet{stg99} detected multiple \HI\ clouds
near Phoenix and \citet{stg99} concluded one of these components was
likely associated with the galaxy. This was confirmed by optical
velocity measurements by \citet{gal01} and \citet{irw02} who both
suggested the \HI\ was blown out, though the loss of \HI\ by
ram-pressure stripping was also possible. Other than the evidence for
these two objects, we have no reason to believe blowouts have occurred
in the galaxies in our sample.

The removal of ISM from a galaxy by a burst of star formation
certainly occurs in some systems. There is ample evidence from
Ly$\alpha$ absorption features in the spectra of distant quasars
\citep{fra82,wan95,cia97} that metals have been ejected from galaxies
into the intergalactic medium. There are also many direct observations
of galaxies currently undergoing this type of event (see
\citealt{mac99} for a recent review). However, there are also numerous
observations that pose serious problems to outflows of this kind {\it
commonly} occurring in dwarfs. \citet{ski95} and \citet{ski97}
describe these observations in detail. The observations presented in
this work propose an additional question: If blowouts are responsible
for the lack of dust/PAHs in low-mass galaxies, how did these blowouts
manage to remove the dust/PAHs {\it without having any impact on the
\HI\ }. Or alternatively, if our sample is unique in that these
galaxies have not experienced blowout, how can our diffuse 8 \micron\
luminosities fit the correlations observed by others so well? Based on
the sum of this evidence we conclude that blowouts cannot solely be
responsible for the lack of diffuse 8 \micron\ emission in the dwarfs
in our sample. Our results are consistent with models by \citet{mac99}
of intermediate- and low-mass galaxies undergoing bursts of star
formation which show that a significant fraction of the gas will be
removed only in galaxies with M$_{gas}$ $\lesssim$ 10$^{6}$ M$_\sun$.
Because this very low gas mass is more consistent with ISM-deficient
`transition galaxies' than with typical dwarf irregulars we do not
believe this process is {\it generally} responsible for the lack of
hot dust/PAH emission in star forming dwarf irregular galaxies.

More recently, in their study of starburst galaxies \citet{oha05} find
a strong anti-correlation between the [\ion{Fe}{2}]/[\ion{Ne}{2}]
ratio (which can be a tracer of supernova activity) and PAH
luminosity. They attribute this anti-correlation to the destruction of
PAH molecules by supernova shocks, and postulate that quiescent (i.e.,
star-forming but non-bursting) dwarf galaxies such as those in our
sample may have stronger PAH emission than in starburst
galaxies. Since this prediction is not confirmed by our observations,
we have considered what physical mechanisms other than the destruction
of PAHs by supernova driven shocks could lead to the strong
anti-correlations between [\ion{Fe}{2}]/[\ion{Ne}{2}] and both PAH
strength and metallicity.  We note that a strong correlation also
exists between [\ion{Fe}{2}]/[\ion{Ne}{2}] and
[\ion{Ne}{3}]/[\ion{Ne}{2}] which is understood simply as the change
in the ionization fraction of Ne in the warm ISM as a function of
metallicity. This correlation can easily explain the two other
correlations observed. However, the destruction of PAHs and dust
grains by supernova shocks could help partly explain the lack of PAH
and hot dust emission in these systems.

An important issue in understanding the diffuse emission at 8 \micron\
is the actual growth site of PAH molecules and dust grains in the
ISM. The idea that dust grains must predominantly be grown through
processes in the ISM rather than in the winds of individual evolved
stars has been discussed for some time \citep{mck89,dra90}. Their
calculations show that in the Galaxy, the mean timescale of a metal
atom in a grain ($\tau_{dest}$) is of order 3$\times$10$^8$ yr, while
the mean `residence time' of a metal atom in the ISM ($\tau_{ISM}$) is
$\sim$10$^9$ yr. The ratio $\tau_{dest}$/($\tau_{dest} + \tau_{ISM}$)
describes the fraction of stardust that would be found in the original
stardust particle (a value of 1 means all of the atoms in a grain have
been there since the creation of the grain in the stellar wind, while
a value of 0 means the grain has been completely rebuilt in the
ISM). Using Galactic values, \citet{dra03} finds a ratio of 0.2,
meaning only 20\% of the atoms will be found in their original
stardust particle.

Accordingly, the question becomes how does this value depend on the
metallicity of the host galaxy? We cannot directly compare the result
of the \citet{dra03} calculation to the galaxies in our sample because
of a lack of simulations on the effects of supernovae and other
energetic processes in low-metallicity dwarf galaxies. This lack of
detailed simulations is quite understandable, given how little is
known about the content of the ISM (especially dust and molecular gas)
in these objects.  However, studies on the evolution of the
multi-phase ISM give us some insight into possible causes of the
drastically different dust content observed between massive spirals
and dwarfs.

Simulations by \citet{nor97} show the formation of molecular gas and
giant molecular clouds is suppressed below metallicities of Z $\sim$
0.03~-~0.1 Z$_{\sun}$ (or roughly 12 + log(O/H) = 7.1-7.7, assuming a
solar value of 12 + log(O/H) = 8.66; \citealt{asp04,mel04}). This
suppression is due to species such as C, O, and CO not being
sufficiently abundant to cool the gas to temperatures necessary to
form molecular hydrogen. Additionally, observations show an abrupt
drop in CO content with metallicity, with no CO having been detected
in any galaxy with 12 + log(O/H) $<$ 7.9 \citep{tay98,ler05}. The lack
of CO in low-metallicity systems is normally attributed to the
inability of low-metallicity systems to self-shield against
dissociating radiation, which shrinks the CO core of a molecular
cloud, but leaves the H$_2$ unaffected \citep{mal88}. This results in
the increase of the CO-to-H$_2$ conversion factor (X$_{CO}$) that has
been observed in Local Group dwarfs \citep{wil95,ver95}.

Together, these studies point to a fundamental change in the phase and
content of the ISM below 12 + log(O/H) $\sim$ 8 (or roughly 20\% of
solar metallicity). It is interesting to note that this is the value
\citet{eng05} find to be the dividing line between systems with PAH
features detected compared to the continuum, and those without. If at
low-metallicity, C, O, and CO are not sufficiently abundant to provide
cooling for molecular cloud formation, and CO is also more easily
dissociated due to the lack of dust to shield it, then the overall
content or filling-factor of molecular gas should decrease in
lower-metallicity systems. Consequently, because grain growth in the
ISM can only take place in cool, dense, molecular clouds
\citep{dra90}, the timescale to regrow grains must be significantly
longer in low-metallicity systems. In contrast, there is no
corresponding change with metallicity in the stellar population, which
is primarily responsible for grain destruction. Therefore, if dust
grains are predominantly grown in the ISM rather than in the winds of
individual stars, we would expect a much lower total dust mass
(compared to the total galaxy mass) at low-metallicity than in the
Galaxy.

Given a long timescale to regrow grains, the lack of diffuse 8
\micron\ emission could be due to the destruction of dust by not just
recent supernovae, but supernovae occurring throughout the evolution
of a galaxy. Future far-IR imaging of the dust distributions of dwarf
galaxies as well as GALEX UV imaging will help elucidate how the
radiation fields and overall dust distributions relate to the diffuse
8 \micron\ emission observed, as well as the evolution of dust
grains. However, IR spectra are also needed to determine whether we
are in fact seeing the emission from hot dust, PAHs, or both.

%%%%%%%%%%%%%%%%%%%%%%%%%%%%%%%%%%%%%%%%%%%%%%%%%%%%%%%%%%%%%%%%%%%%%%%%%

\section{Conclusions}\label{conclusions}
We have shown the spatially resolved distributions of diffuse 8
\micron\ emission for 15 Local Group and other nearby galaxies. We
detect prominent emission from NGC~55, IC~5152, NGC~3109, and IC 1613,
low surface brightness emission from DDO~216, Sextans~A, Sextans~B,
WLM, and in the remaining systems (Antlia, DDO~210, GR~8, Leo~A,
LGS~3, Phoenix, UGCA~438) we observe no diffuse emission.  These data
are the first spatially resolved images of diffuse 8 \micron\ emission
from such low-metallicity objects (12+log(O/H)$\sim$7.5 for Sextans~A
and Sextans~B).

We find that both the nebular metallicities and the strength of the UV
radiation field appear to play important roles in the generation of
this emission. Still, other galaxy properties such as the total dust
content and the fact that the global current star formation rate is
not necessarily a good tracer of the strength of the UV radiation
field in individual star forming regions may complicate
matters. Further, we propose that the lack of emission from hot dust
and PAHs may be due to the combined effects of their destruction by
supernova shocks, coupled with their inability to be regrown in the
ISM.

Finally, our ability to highly resolve these objects at other
wavelengths allows us to rule out some physical causes for the lack of
diffuse 8 \micron\ emission. The presence of a solely young stellar
population that has not had enough time to produce the necessary ISM
components is ruled out by detailed star formation histories which
show that all of the galaxies in our sample formed the bulk of their
stars over 2 Gyr ago. Likewise, blowout of the ambient ISM by vigorous
star formation cannot be responsible for the lack of diffuse 8
\micron\ emission in our sample of galaxies, since \HI\ studies of
these objects show no evidence of this.

%%%%%%%%%%%%%%%%%%%%%%%%%%%%%%%%%%%%%%%%%%%%%%%%%%%%%%%%%%%%%%%%%%%%%%%%%

\acknowledgments D.C.J. wishes to thank Bruce Draine and Terry Jones
for helpful discussions. We also thank the referee for their prompt
and careful reading of the manuscript and their valuable comments.
H.~L. and E.~D.~S. acknowledge partial support from a NASA LTSARP
grant NAG 5-9221 and the University of Minnesota. This work is based
on observations made with the Spitzer Space Telescope, which is
operated by the Jet Propulsion Laboratory, California Institute of
Technology under a contract with NASA. Support for this work was
provided by NASA through contracts 1256406 and 1215746 issued by
JPL/Caltech to R.D.G. at the University of Minnesota.  This research
has made use of NASA's Astrophysics Data System, and of the NASA/IPAC
Extragalactic Database, which is operated by the Jet Propulsion
Laboratory, California Institute of Technology, under contract with
the NASA.

%%%%%%%%%%%%%%%%%%%%%%%%%%%%%%%%%%%%%%%%%%%%%%%%%%%%%%%%%%%%%%%%%%%%%%%%%

\clearpage
\begin{deluxetable}{lccc}
\tablecaption{Galaxy Sample}
\tablewidth{0pt}
\tablehead{
\colhead{Galaxy} & \colhead{Pointing Center ($\alpha$, $\delta$ (J2000))} & \colhead{AOR Key} & \colhead{Mosaic Size}}
\startdata 
Antlia    & 10$^h$04$^m$04$^s$.00, -27$\degr$19$\arcmin$48$\arcsec$.0 & 5053184 & 5$\arcmin$ $\times$ 5$\arcmin$   \\
IC~1613   & 01$^h$04$^m$47$^s$.79,  02$\degr$07$\arcmin$04$\arcsec$.0 & 5051648 & 10$\arcmin$ $\times$ 10$\arcmin$ \\
IC~5152   & 22$^h$02$^m$41$^s$.90, -51$\degr$17$\arcmin$44$\arcsec$.0 & 5055232 & 5$\arcmin$ $\times$ 5$\arcmin$   \\
DDO~210   & 20$^h$46$^m$51$^s$.81, -12$\degr$50$\arcmin$52$\arcsec$.5 & 5054976 & 5$\arcmin$ $\times$ 5$\arcmin$   \\
DDO~216   & 23$^h$28$^m$36$^s$.25,  14$\degr$44$\arcmin$34$\arcsec$.5 & 5055744 & 5$\arcmin$ $\times$ 5$\arcmin$   \\
GR 8      & 12$^h$58$^m$40$^s$.08,  14$\degr$13$\arcmin$00$\arcsec$.4 & 5054464 & 5$\arcmin$ $\times$ 5$\arcmin$   \\
Leo~A     & 09$^h$59$^m$26$^s$.46,  30$\degr$44$\arcmin$47$\arcsec$.0 & 5052416 & 5$\arcmin$ $\times$ 5$\arcmin$   \\
LGS~3     & 01$^h$03$^m$52$^s$.94,  21$\degr$53$\arcmin$05$\arcsec$.0 & 5051392 & 5$\arcmin$ $\times$ 5$\arcmin$   \\
NGC~55    & 00$^h$14$^m$54$^s$.01, -39$\degr$11$\arcmin$49$\arcsec$.3 & 5056000 & 15$\arcmin$ $\times$ 10$\arcmin$ \\
NGC~3109  & 10$^h$03$^m$06$^s$.66, -26$\degr$09$\arcmin$32$\arcsec$.3 & 5052928 & 5$\arcmin$ $\times$ 10$\arcmin$  \\
Phoenix   & 01$^h$51$^m$06$^s$.34, -44$\degr$26$\arcmin$40$\arcsec$.9 & 5052160 & 5$\arcmin$ $\times$ 5$\arcmin$   \\
Sextans~A & 10$^h$11$^m$00$^s$.80, -04$\degr$41$\arcmin$34$\arcsec$.0 & 5053696 & 5$\arcmin$ $\times$ 5$\arcmin$   \\
Sextans~B & 10$^h$00$^m$00$^s$.10,  05$\degr$19$\arcmin$56$\arcsec$.0 & 5052672 & 5$\arcmin$ $\times$ 5$\arcmin$   \\
UGCA~438  & 23$^h$26$^m$27$^s$.52, -32$\degr$23$\arcmin$19$\arcsec$.5 & 5055488 & 5$\arcmin$ $\times$ 5$\arcmin$   \\
WLM       & 00$^h$01$^m$58$^s$.16, -15$\degr$27$\arcmin$39$\arcsec$.3 & 5051136 & 5$\arcmin$ $\times$ 15$\arcmin$  \\
\enddata
\label{sample}
\end{deluxetable}   
\clearpage

%%%%%%%%%%%%%%%%%%%%%%%%%%%%%%%%%%%%%%%%%%%%%%%%%%%%%%%%%%%%%%%%%%%%%%%%%

\clearpage
\begin{deluxetable}{lccccccc}
\tabletypesize{\footnotesize} 
\tablecaption{Galaxy Data}
\tablewidth{0pt}
\tablehead{
\colhead{Galaxy} & \colhead{M$_B$} & \colhead{12 + log(O/H)} & \colhead{Current SFR\tablenotemark{a}} & \colhead{M$_{HI}$ } & \colhead{4.5 \micron\ flux\tablenotemark{b}} & \colhead{Diffuse 8.0 \micron\ flux} & Scale \\ 
\colhead{} & \colhead{(mag)} &\colhead{(dex)} &\colhead{(M$_\sun$/yr)} &\colhead{(M$_\sun$)} & (mJy) & (mJy) & factor}
\startdata 
Antlia    & -10.2 & -    & -      & 9.7$\times$10$^5$ &  1.86 $\pm$ 0.2	 & 3.3  $\pm$ 1.5 & 0.4 \\
DDO~210   & -9.9  & -    & -      & 1.9$\times$10$^6$ &  2.14 $\pm$ 0.2	 & 3.4  $\pm$ 1.7 & 0.4 \\ 
DDO~216   & -12.3 & 7.93 & 0.0    & 5.4$\times$10$^6$ &  30.5 $\pm$ 3.1	 & 11   $\pm$ 1.0 & 0.4 \\
GR8       & -11.2 & 7.65 & 0.0007 & 4.5$\times$10$^6$ &  2.95 $\pm$ 0.30 & 6.2  $\pm$ 1.0 & 0.4 \\
IC~1613   & -14.2 & 7.62 & 0.003  & 5.4$\times$10$^7$ & 91.53 $\pm$ 9.2  & 25   $\pm$ 8.0 & 0.4 \\
IC~5152   & -14.5 & 7.92 & -      & 5.9$\times$10$^7$ &   14  $\pm$ 15	 & 160  $\pm$ 25  & 0.4 \\
Leo~A     & -11.3 & 7.38 & 0.0003 & 8.0$\times$10$^7$ &  8.65 $\pm$ 0.87 & 0.9  $\pm$ 1.0 & 0.4 \\
LGS~3     & -9.9  & -    & 0.0    & 4.2$\times$10$^5$ &  0.37 $\pm$ 0.04 & 8.7  $\pm$ 3.4 & 0.4 \\
NGC~55    & -17.5 & 8.05 & 0.18   & 1.4$\times$10$^9$ &   785 $\pm$ 79	 & 1400 $\pm$ 310 & 0.7 \\
NGC~3109  & -15.2 & 8.06 & 0.02   & 6.9$\times$10$^8$ &   131 $\pm$ 13	 & 55   $\pm$ 11  & 0.4 \\
Phoenix   & -9.5  & -    & -      & 2.0$\times$10$^5$ &  7.31 $\pm$ 0.7  & 2.7  $\pm$ 1.7 & 0.4 \\
Sextans~A & -14.2 & 7.54 & 0.002  & 7.8$\times$10$^7$ &  8.39 $\pm$ 0.84 & 0.3  $\pm$ 0.1\tablenotemark{c} & 0.4 \\
Sextans~B & -13.8 & 7.53 & 0.0008 & 4.5$\times$10$^7$ &  23.0 $\pm$ 2.3	 & 8.3  $\pm$ 3.3 & 0.4 \\
UGCA~438  & -11.7 & -    & -      & 6.2$\times$10$^6$ &  5.0  $\pm$ 0.5	 & 7.0  $\pm$ 2.7 & 0.4 \\
WLM       & -14.5 & 7.83 & 0.003  & 6.1$\times$10$^7$ &  45.8 $\pm$ 4.6	 & 15   $\pm$ 6.3 & 0.4 \\
\enddata
\label{data}
\tablenotetext{a}{The systematic uncertainty in the current star
formation rates derived from H$\alpha$ luminosities is estimated by
\citet{ken83} to be roughly 50\% or less.}  
\tablenotetext{b}{4.5 \micron\ fluxes are from \citet{lee06} except
Antlia, DDO~210, Leo~A, Phoenix, and UGCA~438. Fluxes for these
targets were calculated as described in \citet{lee06}.}
\tablenotetext{c}{The
diffuse 8 \micron\ flux listed for Sextans~A is only for the large
\HII\ region in the southeast. See \S \ref{sextans_a}.}
\tablerefs{All M$_B$, Current SFR (based on H$\alpha$ fluxes), and
M$_{HI}$ values from \citet{mat98}. 12~+~log(O/H) values are from
\citet{lee06}. Scale factor refers to the multiplicative value applied
to the channel 2 image for continuum subtraction (see \S
\ref{observations}).}

\end{deluxetable}
\clearpage
%%%%%%%%%%%%%%%%%%%%%%%%%%%%%%%%%%%%%%%%%%%%%%%%%%%%%%%%%%%%%%%%%%%%%

\begin{figure}
\begin{center}
%\plotone{f1.eps}
\caption{\label{n55}Approximately 5$\arcmin$$\times$11$\arcmin$ images of NGC~55 
at 4.5 \micron\ (top) and continuum subtracted 8 \micron\ (bottom).}
\end{center}
\end{figure}

\begin{figure}
\begin{center}
\epsscale{0.85}
%\plotone{f2.eps}
\caption{\label{3detect} 4.5 and diffuse 8 \micron\ images of IC 5152
(top), NGC~3109 (middle), and IC~1613 (bottom). For reference, the
white box in the diffuse 8 \micron\ image of IC~1613 is the same as
Frame L in the H$\alpha$ study of \citet{hod90}.}
\end{center}
\end{figure}

\begin{figure}
\begin{center}
%\plotone{f3.eps}
\caption{\label{split1} Top: 4.5 and diffuse 8 \micron\ images of WLM.
Bottom panel: 4.5 and 8 \micron\ image of Sextans A.  The 8
\micron\ image of Sextans A is not continuum-subtracted.  In both
panels the white contours represent the H$\alpha$ surface brightness
distribution \citep{mas02}.}
\end{center}
\end{figure}

\begin{figure}
\begin{center}
%\plotone{f4.eps}
\caption{\label{split2} 4.5 and diffuse 8 \micron\ images of
Sextans~B (top) and DDO~216 (bottom). The white contours represent the H$\alpha$
surface brightness distribution (H$\alpha$ images from \citet{mas02}). }
\end{center}
\end{figure}

\begin{figure}
\begin{center}
\epsscale{0.7}
%\plotone{f5.eps}
\caption{\label{nodetect} 4.5 \micron\ images of Antlia, DDO~210, GR8, 
Leo~A, LGS~3, Phoenix, and UGCA~438.}
\end{center}
\end{figure}

\begin{figure}
\begin{center}
\epsscale{1}
%\plotone{f6.eps}
\caption{\label{lumin} Diffuse 8 \micron\ luminosity normalized by the
total stellar mass as a function of (a) the nebular oxygen abundance
and (b) log of the current star formation rate (see Table \ref{data}
for details). Our data are shown (filled squares) along with data from
\citet{eng05} (stars), and \citet{ros06} (open circles). The total
stellar mass was inferred from the 4.5 \micron\ luminosity as
described in \citet{lee06} assuming a Salpeter IMF, and the adopted
distances to the galaxies in our sample are from \citet{mat98}. The
data of \citet{eng05} are only shown in part (a) because they use a
different tracer of current star formation activity. One sigma error
bars are shown where available except the abundance error on the
\citet{eng05} data, which they adopt to be 0.05 dex. Errors in
log(L/M) only include diffuse 8 \micron\ flux errors; no error is
assumed on the stellar mass estimate. }
\end{center}
\end{figure}

\begin{figure}
\begin{center}
\epsscale{1}
%\plotone{f7.eps}
\caption{\label{3corr}Absolute diffuse 8 \micron\ magnitude versus (a)
absolute B magnitude, (b) nebular oxygen abundance, and (c) current
star formation rate (see Table \ref{data} for details). One sigma
error bars are shown where available, and Sextans~A is plotted as a
lower limit (see \S \ref{sextans_a}).}
\end{center}
\end{figure}

\begin{figure}
\begin{center}
%\plotone{f8.eps}
\caption{\label{3groups}Absolute diffuse channel 4 magnitude vs oxygen
abundance (left) and log of the current star formation rate (right)
(see Table \ref{data} for details). Our data are shown (filled
squares) along with data from \citet{eng05} (stars), and \citet{ros06}
(open circles). The data of \citet{eng05} are only shown in the left
panel because they use a different tracer of current star formation
activity. One sigma error bars are shown where available, except the
abundance error on the \citet{eng05} data, which they adopt to be 0.05
dex. Sextans A is shown as a lower limit (see \S \ref{sextans_a}).}
\end{center}
\end{figure}

\end{document}